\documentclass[twocolumn,aps, prl, groupedaddress,epsf]{revtex4}

\usepackage{graphicx}

\begin{document}

\title{Scaling of Resistance and Electron Mean Free Path of Single-Walled Carbon Nanotubes}

\author{Meninder Purewal$^1$}
\author{Byung Hee Hong$^2$}
\author{Anirudhh Ravi$^2$}
\author{Bhupesh Chandra$^3$}
\author{James Hone$^3$}
\author{Philip Kim$^2$}

\affiliation{$^1$ Department of Applied Physics, Columbia
University, New York, New York 10027} \affiliation{$^2$ Department
of Physics, Columbia University, New York, New York 10027}
\affiliation{$^3$ Department of Mechanical Engineering, Columbia
University, New York, New York 10027}

\begin{abstract}
We present an experimental investigation on the scaling of
resistance in individual single walled carbon nanotube devices with
channel lengths that vary four orders of magnitude on the same
sample. The electron mean free path is obtained from the linear
scaling of resistance with length at various temperatures. The low
temperature mean free path is determined by impurity scattering,
while at high temperature the mean free path decreases with
increasing temperature, indicating that it is limited by
electron-phonon scattering. An unusually long mean free path at room
temperature has been experimentally confirmed. Exponentially
increasing resistance with length at extremely long length scales
suggests anomalous localization effects.

\end{abstract}

\maketitle


Single walled carbon nanotubes (SWNTs) are 1D conductors that
exhibit a rich variety of low dimensional charge transport
phenomena~\cite{Dresselhaus1}, including ballistic
conduction~\cite{Kong2, Liang3, Mann4, Javey5, Javey6},
localization~\cite{Navarro7} and 1D variable range
hopping~\cite{Gao8}.  The electron mean free path, $L_{m}$, is one
of the important length scales that characterize the different 1D
transport regimes.  One method of determining $L_{m}$ in SWNTs is to
measure ballistic conduction for a given device channel length.
However, this method yields a lower bound of $L_{m}$, and works only
at low temperature ~\cite{Kong2, Liang3, Mann4, Javey5} or at higher
temperature for small length scales ($<$60~nm)~\cite{Javey6}.
Another approach to obtain $L_{m}$ at room temperature is to employ
scanning probe microscopy to measure the linear scaling of the
channel resistance~\cite{Park9}, or use non-invasive multi-terminal
measurements~\cite{Gao10}.  Due to the experimental limitations of
these approaches, the characterization of $L_{m}$ for the same SWNTs
over a range of temperatures is yet to be realized.

Recent advances in the growth of extremely long SWNTs ($>$1~mm)~\cite{Hong11} now allow
for an intensive study on their intrinsic properties.  In this letter, we present
experimental measurements on the scaling behavior of resistance in individual,
millimeter long SWNTs for the temperature range of 1.6~-~300~K. From the linear scaling
of resistance, the temperature dependent electron mean free path is calculated for
each temperature. Beyond the linear scaling regime, we observe that the resistance
increases exponentially with length, indicating localization behavior.

\begin{figure}
\includegraphics[width=85mm]{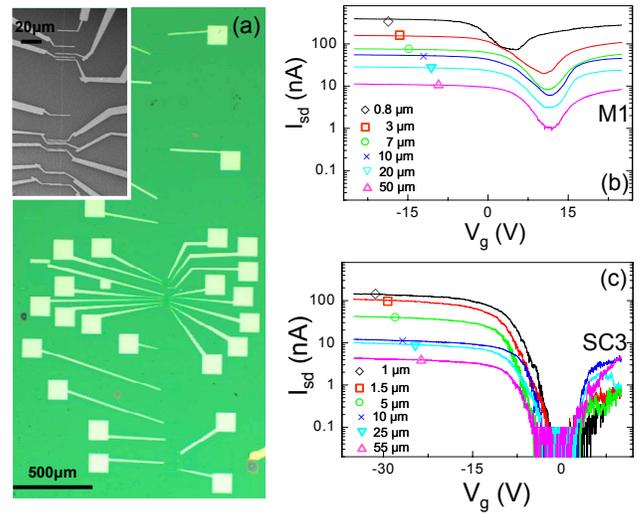}
\caption{(a)Optical image showing typical SWNT devices
with multiple Pd electrodes. (Inset) Scanning electron microscope
image of an isolated SWNT contacted with these electrodes. Room
temperature $I_{SD}(V_{g})$ of selected channel lengths for (b)
metallic SWNT (M1) and (c) semiconducting SWNT (SC3) with
$V_{SD}=$~6.4 and 2.7~mV, respectively.} \label{fig1}
\end{figure}

Macroscopically long and straight individual SWNTs were grown on a
degenerately doped Si/SiO$_{2}$ substrate ($t_{ox}=$~500~nm) using
the chemical vapor deposition method described in Ref.\cite{Hong11}.
This was followed by the fabrication of multiple Pd electrodes with
various separations (200~nm- 400~$\mu$m)(Fig.~\ref{fig1}(a)). Pd
electrodes were chosen to create highly transparent SWNT-electrode
contacts~\cite{Mann4}. The diameters of the SWNTs were measured by
atomic force microscope (AFM). We chose SWNTs with diameter
 $d$ less than 2.5 nm to exclude any possibility of including multiwalled nanotubes (MWNT)
 in this study. In addition, we confirmed that the high bias saturation current is
 $<$~30~$\mu$A for all SWNTs studied~\cite{Yao12}, assuring that the samples consisted
 of single tubes rather than small bundles or MWNTs. The substrate was used as a gate
 electrode to tune the chemical potential of the sample by the application of a gate
 voltage $V_{g}$.  A small dc source-drain bias voltage ($<$~10~mV), $V_{SD}$, was applied
 between pairs of consecutive electrodes, and the two-terminal linear response conductance
 was determined from the measured source-drain current $I_{SD}$.

Fig.~\ref{fig1}(b-c) shows the measured $I_{SD}$ as a function of
$V_{SD}$ for selected channel length sections on two
representative SWNTs. All curves exhibit a `gap' like feature - a
range of $V_{g}$ where $I_{SD}$ is suppressed. On the same SWNT,
every device (pair of consecutive electrodes) shows a similar
$I_{SD}(V_{g})$ up to a length-dependent multiplicative factor,
once we align the centers of the gap region for each curve. The
similarity of the $I_{SD}(V_{g})$ behavior in different sections
for each SWNT sample indicates that the corresponding `gap'
features are derived from the intrinsic electronic structure of
the SWNT rather than the effects of random local variation.

\squeezetable
\begin{table*}
\caption{\label{tab1}Device characteristics for SWNTs used in this
study.  The character M (SC) is designated for metallic
(semiconducting) SWNTs.}
\begin{ruledtabular}
\begin{tabular}{c|c|c|c|c|c|c|c|c|c|c|c|c|c}
   & M1 &M2 &M3 &M4 &SC1 &SC2 &SC3 &SC4 &SC5 &SC6 &SC7 \\ \hline
 $d(nm)$& $2.0\pm.2$& $1.3\pm.4$& $1.7\pm.6$ & $1.6\pm.4$& $1.6\pm.4$& $1.8\pm.6$& $1.9\pm.4$&
 $2.1\pm.2$& $2.2\pm.2$& $2.0\pm.6$& $2.2\pm.2$\\
 $R_{c}(k\Omega)$& $7.9\pm.8$& $11.5\pm2.9$ & $8.3\pm2.5 $& $12.0\pm4.4 $& $10.2\pm4.5$&
 $14.9\pm5.7$& $10.4\pm.9$ & $7.0\pm2.3$ & $25.4\pm4.2$ & $6.9\pm40$& $21.8\pm14$\\
 $\rho_{sat}(k\Omega/\mu$m) & $0.76\pm.02$ & $ 0.87\pm.02 $ & $0.93\pm.01$ & $6.5\pm.08 $ &
 $2.95\pm.05$ & $3.61\pm.05$ & $4.64\pm.01$ & $5.91\pm.12$ & $8.13\pm.31 $ &
 $14.1\pm.19 $ & $16.3\pm.13$\\
 $L_{m}^{sat}(\mu$m) & $8.56\pm.23$ & $7.65\pm.17$ & $7.07\pm.08$ & $1.00\pm.01$ &
 $2.24\pm.04$ & $1.83\pm.03$& $1.40\pm.01$& $1.10\pm.02$ & $0.80\pm.03$ &
 $0.47\pm.01$ & $0.40\pm.01$\\
\end{tabular}
\end{ruledtabular}
\end{table*}

We use the qualitatively different $I_{SD}(V_{g})$ behaviors of
different SWNTs to categorize them as metallic (M-NT) or
semiconducting nanotubes (S-NT). Typical S-NTs
(Fig.~\ref{fig1}(c)) exhibit an off current region
$I_{SD}<10^{-10}$A when the Fermi energy $E_{F}$ lies n the energy
gap ~\cite{Tans13, Appenzeller14}. On the other hand, a weaker
suppression of $I_{SD}(V_{g})$ is observed in the `small gap'
region in M-NTs (Fig.~\ref{fig1}(b)). The `small gap' in M-NTs has
been attributed to the curvature-induced energy gap
$E_{g}<$100~meV~\cite{Zhou15}, which is distinguished from the
S-NT energy gap, which scales with diameter as
$E_{g}\sim$~1/$d$~(nm)~\cite{Dresselhaus1}. Among the 11 SWNTs we
studied in this letter, we found 4 M-NTs and 7 S-NTs. Each of
these SWNTs exhibit a gap centered at $V_{g}>0$, indicating their
$p$-doped nature. At large negative gate voltage ($V_{g}<-20$~V),
$E_{F}$ lies well outside of the gap region and $I_{SD}(V_{g})$
saturates to $I_{SD}^{sat}$, whose value depends only on the
applied $V_{SD}$ and channel length $L$ of the SWNT section. The
two-terminal resistance of the SWNT section is then obtained from
$R(L)= V_{SD}/I_{SD}^{sat}$. We note that four-terminal resistance
measurements are possible for each section by utilizing the
available multiple electrode configuration. However, in our
experiment, the four terminal measurements yield essentially
similar results to the two terminal $R(L)$, which prevents
separation of the `contact' resistance contribution from $R(L)$.
Such inseparable contact resistance between SWNT-metal electrodes
was reported to be caused by the invasiveness of metal
contacts~\cite{Bezryadin16}.

\begin{figure}
\includegraphics[width=80mm]{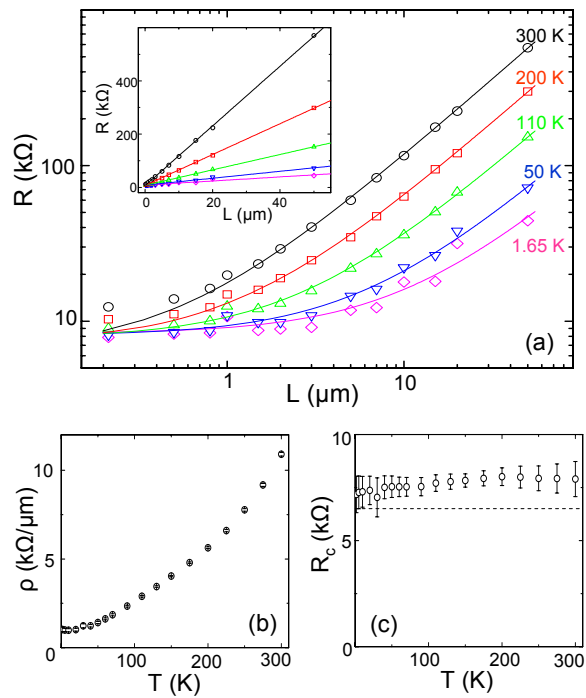}
\caption{(a) (Inset) $R(L)$ for sample M1 at select temperatures
ranging from 1.6~-~300~K. (Main) A log-log plot highlights the
behaviors at different lengths scaling 3 orders of magnitude. From
the linear fits (solid lines) of these data points, we obtain the
1D resistivity (b) and the contact resistance (c) at different
temperatures.  The dashed line in (c) represents $R_{Q}$.}
\label{fig2}
\end{figure}

We designed many pairs of electrodes with different $L$ on each
SWNT so that the scaling of $R(L)$ can be studied for a specific
sample at a given temperature $T$. Fig.~\ref{fig2}(a) show $R(L)$
of a representative SWNT measured in the temperature range of
1.6~-~300~K and with an $L$ range of 200~nm~-~50~$\mu$m. In these
ranges, $R(L)$ increases linearly and appears to converge to a
finite value for small $L$ (inset to Fig.~\ref{fig2}(a)). We found
that this scaling behavior can be described well by a simple
linear dependence with an offset: $R(L)=\rho L+R_{c}$, where
$\rho$ and $R_{c}$ are interpreted as the 1D resistivity and
contact resistance, respectively. The solid lines in
Fig.~\ref{fig2}(a) are the two parameter line fits of the data
points at a given $T$ value. From these fits, $R_{c}(T)$ and
$\rho(T)$ are obtained as shown in Fig.~\ref{fig2}(b) and
Fig.~\ref{fig2}(c), respectively. For this sample, $R_{c}$ remains
fairly constant at $\sim$8~k$\Omega$ and $\rho(T)$ exhibits
typical metallic behavior, i.e. it decreases with $T$ and
saturates to a value $\rho_{sat}$ at low temperatures. Similar
scaling behavior of $R(L)$ is observed in other SWNTs, from which
both $R_{c}$ and $\rho(T)$ are extracted within the linear scaling
regime. Table ~\ref{tab1} summarizes $d$, $R_{c}$, and
$\rho_{sat}$ for the 4 M-NTs and 7 S-NTs considered in this study.
To understand the scaling of $R(L)$ in Fig.~\ref{fig2}, we begin
with the two-terminal Landauer-Buttiker formula applied to
SWNTs~\cite{Park9}. If we consider 4 low-energy channels in the
SWNT, 2 each for spin and band degeneracy, then the scaling of
resistance is given by $R(L)=(h/4e^{2})(L/L_{m}+1)+R_{nc}$, where
$e$ and $h$ are electron charge and Plank constant and $L_{m}$ and
$R_{nc}$ are the electron mean free path and the non-transparent
contact resistance, respectively. Note that we separate out the
contribution of $R_{nc}$ from the total contact resistance
$R_{c}$, so that the contact resistance becomes the quantum
resistance $R_{Q}=h/4e^{2}$ when the contacts become fully
transparent. From the experimentally obtained $\rho(T)$ and
$R_{c}$, we can deduce $L_{m}=R_{Q}/\rho(T)$ and $R_{nc}=R_{c} -
R_{Q}$ for each of our SWNT samples. In particular, we note that
$R_{nc}\lesssim R_{Q}$ for the majority of our samples, suggesting
that the barrier at the contacts is very thin and adds only a
negligible contribution when $L$ becomes substantially large.

\begin{figure}
\includegraphics[width=80mm]{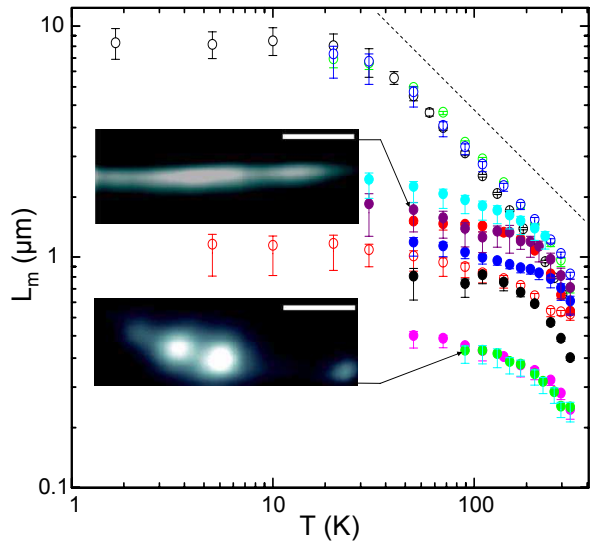}
\caption{(color online) The electron mean free path for the
samples listed in Table~\ref{tab1} at different temperatures.
Most metallic SWNTs (open circles) saturate at higher values than
that of semiconductors (closed circles). The dashed line
represents $T^{-1}$ dependence. The insets show scanning gate
microscopy images taken on devices SC2 (upper) and SC7 (lower).
Lighter color corresponds less current in the SWNT. The defects in
the SWNT are highlighted by the bright region (suppressed current)
on the SWNT.  The scale bar is 500nm.} \label{fig3}
\end{figure}

We now discuss the temperature dependent behavior of the mean free
path. Fig.~\ref{fig3} is the central result of this letter,
showing $L_{m}(T)$ of the SWNTs listed in Table~\ref{tab1}.
Overall, $L_{m}(T)$ exhibits different behaviors in two regimes
separated by $T_{cr}$: (i) the high temperature regime
($T>T_{cr}$) where $L_{m}\sim T^{-1}$ (dashed line in
Fig.~\ref{fig3}), which indicates that inelastic scattering
between electrons and acoustic phonons is dominant~\cite{Park9,
Avouris17} regardless of chirality~\cite{Zhou17}; and (ii) the low
temperature regime $(T<T_{cr})$ where $L_{m}$ saturates to the the
tube specific $L_{m}^{sat}$. In this low temperature limit, the
phonons freeze out and $L_{m}^{sat}$ is determined by the
temperature independent elastic scattering with impurities. We
believe the widely spread $L_{m}^{sat}$ values (0.4-10$~\mu$m) in
(ii) are a result of each SWNT sample having a static disorder of
different strengths and densities. We employ scanning gate
microscopy (SGM)~\cite{Bachtold18} to image this static disorder.
Indeed, the SGM images on S-NTs (insets to Fig.~\ref{fig3}) reveal
that the SWNT with a shorter $L_{m}^{sat}$ shows more defects.
Note also that we have experimentally confirmed that $L_{m}$ is
generally much higher for M-NTs than that of S-NTs. This is an
indication that the scattering of electrons is strongly suppressed
in M-NTs, as predicted by Ando et al.~\cite{Ando19} and McEuen et
al.~\cite{McEuen20}. In M-NTs we have experimentally shown that
the ballistic electron conduction is possible for channel lengths
up to 8~$\mu$m at low temperature and 0.8~$\mu$m even at room
temperature.

\begin{figure}
\includegraphics[width=80mm]{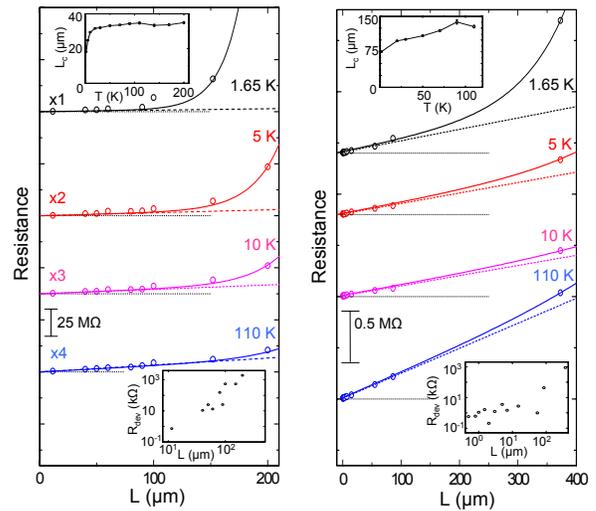}
\caption{$R(L)$ in the non-linear regimes for samples (a) SC6
($L_{m}^{sat}\approx$ 460~nm) and (b)M3 ($L_{m}^{sat}\approx
7~\mu$m). Note that the data is magnified in (a) for clarity. The
dashed line is an extension of the linear regime and the solid
line is a fit for all data. $R_{dev}$ shows the absolute value of
the difference between the actual device resistance and the
corresponding linear resistance at 110~K (lower inset a) and
1.65~K (lower inset b).The non-linearity increases with decreasing
temperature, which is reflected in the value of $L_{c}$(upper
insets).} \label{fig4}
\end{figure}

Finally, we turn our attention to the non-linear scaling of
$R(L)$. Fig.~\ref{fig4} presents $R(L)$ beyond the linear scaling
regime of a representative S-NT and M-NT. At extremely long length
scales and low temperatures, $R(L)$ deviates from the linear
dependence extended from the linear regime (dashed lines in main
figure and see also $R_{dev}=R(L)-R_{c}-R_{Q}L/L_{m}$ in lower
insets). Since $R(L<<L_{m}^{sat})\sim R_{Q}$ for all temperatures,
we emphasize here that this non-linear behavior in $R(L)$ is
solely due to increasing electron scattering in the bulk part of
the SWNTs rather than an increasing barrier between the SWNT and
electrodes. In order to experimentally determine the critical
length scale $L_{c}$ beyond which the non-linear behaviors is
dominant, we use a phenomenological equation:
$R(L)=R_{c}+R_{Q}(L/L_{m}+e^{L/L_{c}})$ to fit the data (solid
curves in Fig.~\ref{fig4}). While $L_{c}$ shows a strong sample
dependent behavior, generally we found $L_{c}>>L_{m}$ in all
temperature ranges, with the temperature dependence exhibiting a
trend of increasing $L_{c}$ with increasing $T$(upper insets to
Fig.~\ref{fig4}). This observed behavior of $L_{c}(T)$ excludes
the quantum interference related to strong localization effects
such as Anderson Localization~\cite{Navarro7} from the possible
scenarios. In particular, in the high temperature regime
$(T>T_{cr})$, the phase coherence length $L_{\phi}$ is limited by
the phase-breaking electron-phonon scattering, and thus
$L_{\phi}\sim L_{m} << L_{c}$, inviting further study to elucidate
the observed localization behavior beyond the strong localization
limit~\cite{Roche21, Roche22}.

In conclusion, we determine the length dependent resistance for
SWNTs with channel lengths ranged 200~nm - 400$~\mu$m. From the
scaling behavior we evaluate the electron mean free path and
localization length of the SWNT for a range of temperatures. While
the low temperature mean free path is determined by the impurity
scattering, an unusually long mean free path is demonstrated at room
temperature, even with the dominant electron-phonon scattering.

We thank I. Aleiner, B. Altshuler, and P. Jarillo-Herrero for
helpful discussions. This work is supported by the NSF NIRT(ECS
0507111), CAREER (DMR-0349232), NSEC (CHE-0117752), and the New York
State Office of Science, Technology, and Academic Research (NYSTAR).


\begin{references}

\bibitem{Dresselhaus1}R. Saito, G. Dresselhaus, and M.S. Dresselhaus,
Physical Properties of Carbon Nanotubes (Imperial College Press, London 1998).

\bibitem{Kong2}J. Kong, E. Yenilmez, T.W. Tombler, W. Kim, H. Dai,
R.B. Laughlin, L. Liu, C.S. Jayanthi, and S.Y. Wu, Phys. Rev.
Lett. \textbf{87}, 106801 (2001).

\bibitem{Liang3}W. Liang, M. Bockrath, D. Bozovic, J.H. Hafner,
M. Tinkham, and H. Park, Nature \textbf{411}, 665 (2001).

\bibitem{Mann4} D. Mann, A. Javey, J. Kong,
Q. Wang, and H. Dai, Nano Lett. \textbf{3}, 1541 (2003).

\bibitem{Javey5} A. Javey, J. Guo, Q. Wang,
M. Lundstrom and H. Dai, Nature, \textbf{424}, 654 (2003).

\bibitem{Javey6} A. Javey, J. Guo, M. Paulsson, Q. Wang,
D. Mann, M. Lundstrom, and H. Dai, Phys. Rev. Lett. \textbf{92},
106804 (2004).

\bibitem{Navarro7} C. Gomez-Navarro, P.J. de Pablo, J. Gomez-Herrero,
B. Biel, F.J. Garcia-Vidal, A. Rubio and F. Flores, Nat. Mater.
\textbf{4}, 534 (2005).

\bibitem{Gao8} B. Gao, D.C. Glattli, B. Placais and A. Bachtold,
Phys. Rev. B \textbf{74}, 085410 (2006).

\bibitem{Park9} J. Park, S. Rosenblatt, Y. Yaish, V. Sazonova, H. Ustunel,
S. Braig, T.A. Arias, P.W. Brouwer and P.L. McEuen, Nano Lett.
\textbf{4}, 517 (2004).

\bibitem{Gao10} B. Gao, Y.F. Chen, M.S. Fuhrer, D.C. Glattli, and
A. Bachtold, Phys. Rev. Lett. \textbf{95}, 196802 (2005).

\bibitem{Hong11}B.H. Hong, J.Y. Lee, T. Beetz,
Y. Zhu, P. Kim, and K.S. Kim, J. Am. Chem. Soc. \textbf{127},
15336 (2005).

\bibitem{Yao12} Z. Yao, C.L. Kane, and C. Dekker,
Phys. Rev. Lett. \textbf{84}, 2941 (2000).

\bibitem{Tans13} S.J. Tans, A.R.M. Verschueren, and C. Dekker, Nature \textbf{393}, 49 (1998).

\bibitem{Appenzeller14} J. Appenzeller, J. Knoch, V. Derycke, R. Martel,
S. Wind and Ph. Avouris, Phys. Rev. Lett. \textbf{89}, 126801
(2002).

\bibitem{Zhou15} C. Zhou, J. Kong, and H. Dai, Phys. Rev. Lett. \textbf{84}, 5604 (2000).

\bibitem{Bezryadin16}A. Bezryadin, A. R. M. Verschueren,
S. J. Tans, and C. Dekker, Phys. Rev. Lett. \textbf{80}, 4036
(1998).

\bibitem{Avouris17} V. Perebeinos, J. Tersoff, and Ph. Avouris, Phys. Rev. Lett. \textbf{94}, 086802 (2005).

\bibitem{Zhou17} X. Zhou, J. Park, S. Huang, J. Liu,
and P. L. McEuen, Phys. Rev. Lett. \textbf{95}, 146805 (2005).

\bibitem{Bachtold18} A. Bachtold, M.S. Fuhrer, S. Plyasunov,
M. Forero, E.H. Anderson, A. Zettl, and P.L. McEuen, Phys. Rev.
Lett. \textbf{84}, 6082 (2000).

\bibitem{Ando19} T. Ando and T. Nakanishi, Jpn. J. Appl. Phys. \textbf{67}, 1704 (1998).

\bibitem{McEuen20} P.L. McEuen, M. Bockrath, D.H. Cobden,
Y. G. Yoon, and S.G. Louie, Phys. Rev. Lett. \textbf{83}, 5098
(1999).

\bibitem{Roche21} F. Triozon, S. Roche, A. Rubio, and D. Mayou, Phys. Rev. B \textbf{69},
121410(R) (2004).

\bibitem{Roche22} R. Avriller, S. Latil, F. Triozon, X. Balse, and S. Roche, Phys. Rev. B \textbf{74},
121406(R) (2006).

\end{references}
\end{document}